\documentclass[preprint, showpacs,floatfix,pre]{revtex4}

\usepackage{graphicx}
\usepackage{dcolumn}
\usepackage{bm}

\begin{document}

\title{Phase Behavior of Wormlike Rods}
\author{Giorgio Cinacchi}
\altaffiliation{Present affiliation: ARCES, Universit\`{a} di Bologna,
Via Vincenzo Toffano 2, I--40125 Bologna, Italy;
Present contact address: Scuola Normale Superiore, Piazza dei
Cavalieri 7, I-56126 Pisa, Italy}
\email{g.cinacchi@sns.it}
\author{Luca De Gaetani}
\email{degaetani@dcci.unipi.it}
\affiliation{
Dipartimento di Chimica, Universit\`{a} di Pisa, \\
Via Risorgimento 35, I--56126 Pisa, Italy
}

\date{\today}

\begin{abstract}
\noindent
By employing Molecular Dynamics computer simulations, the
phase behavior   of systems
of rodlike particles with varying degree of internal flexibility  
has been traced from the perfectly rigid rod limit
till very flexible particles,
and from the high density region till the 
isotropic phase.
From the perfectly rigid rod limit and  enhancing the internal flexibility,
the range of the smectic A phase is squeezed out by the concomitant action
of the scarcely affected crystalline phase
at higher density and the nematic phase at lower
density, until it disappears.  These results confirm
the supposition, drawn from previous theoretical, simulational
and experimental studies, that the smectic A phase is destabilized
by introducing and enhancing the degree of particle internal
flexibility. However, no significant changes in the order of nematic--to--smectic A
phase transition, which appears always first order, nor in the value of
the layer spacing, are observed upon varying the degree of
particle internal flexibility. Moreover, no evidence of a columnar
phase, which was tought of as a possible superseder of the smectic A phase in flexible rods,
has been obtained.

\end{abstract}


\maketitle

Particles with only steric, repulsive interactions are basic models
with which to study
the phase behavior and properties of condensed matter systems.
If the particles are anisometric, a rich phase behavior may be observed,
with the crystalline  and isotropic fluid phases possibly bracketing 
phases of intermediate order. 
Hard spherocylinders \cite{Onsager} provide an emblematic example. They were shown 
to exhibit first  smectic A (S$_{\rm{A}}$), and then also  nematic (N), 
liquid--crystal phases 
upon increasing the aspect ratio\cite{BolhuisFrenkel}.
Hard spherocylinders and related models are pertinent to 
a host of experimental systems where lyotropic liquid crystals
take place. They include systems of biological, organic and inorganic
origin. Examples are TMV and $fd$ viruses \cite{fraden}, DNA \cite{livolant}, 
$\beta$--FeOOH particles \cite{maeda} and   rod--shaped Carbon
nanotube \cite{nanotubiC} and  nanocrystals \cite{jacs}.
For many of these systems, however, rigid models provide only a first
good  approximation. In fact, most of the particles constituting the experimental
suspensions
are, to a certain extent, flexible. 
Internal flexibility is expected to influence
the self--assembly characteristics
of the rodlike particles.
The comprehension of self--assembly  mechanisms is important
\emph{per se} but also because, as in the case of nanorods, their strict 
relationship 
with the properties of the sample may contribute to a successful  exploitation
of the materials. 

Here, the study  of a  completely rigid rodlike model,
performed  earlier \cite{prenostro},
has been extended to investigate the effect of internal flexibility on the phase behavior
of rodlike systems. 

To this end,  systems of elongated particles have been simulated with the Molecular
Dynamics (MD) technique \cite{allentilde}.
The particles, of mass $m$,  are wormlike  and formed by nine beads.
Within a particle, contiguous beads are kept at a fixed distance of
0.6$\sigma$,  $\sigma$ being the quantity
defining the scale of lengths, 
while a harmonic bending interaction exists between three contiguous beads, $l$ and $n$,
 $m$: 
\begin{equation}
v_{lmn}(\theta)=\frac{1}{2}K\left(\theta-\pi \right)^2.
\label{armonico}
\end{equation}
In the equation above, $v_{lmn}$ is the angular potential energy, 
$\theta$ is the angle formed by the two relevant bonds
 and  $K$ the force constant,
regulating the degree of internal flexibility.
In addition, between two non--contiguous beads a repulsive interaction
exists of the following form:
\begin{equation}
u_{ij}\left(r\right)=  \left\{
\begin{array}{l}
4\epsilon\left[\left(\frac{\sigma}{r}\right)^{12}-
\left(\frac{\sigma}{r}\right)^6+\frac{1}{4}\right], r \leq 2^{\frac{1}{6}} \sigma \\
\\
0, r > 2^{\frac{1}{6}} \sigma
\end{array}
\right.
\label{defpot}
\end{equation}
In the equation above, $u_{ij}$ is the interaction potential energy 
between beads $i$ and $j$, separated by a distance $r$, while
$\epsilon$ is the  the quantity   defining the scale of
energies.   
Equation \ref{defpot} describes  also the interaction potential energy 
between any two
beads belonging to two different particles, so that the the interaction potential
energy between the wormlike rods $I$ and $J$ is given by:
\begin{equation}
U_{IJ} = \sum_{k_{I}=1}^{9} \sum_{k_{J}=1}^9 u_{k_{I}k_{J}}.
\label{defpot2}
\end{equation}

Systems of N=600
wormlike particles have been simulated for several values of $K$, 
covering six order of magnitude of the force constant, ranging 
from perfectly rigid to very flexible rods.
The computations have been performed at a fixed pressure of $P^*$=24.716
$\frac{\sigma^3}{\epsilon}$
and varying the temperature $T^*=$$k_BT/\epsilon$, with $k_B$ the 
Boltzmann constant. 
Pressure  and temperature have been maintained at the
pre--selected values with either the Nos\'{e}--Hoover thermostat, coupled with
the Parrinello--Rahman barostat or the weak coupling method  \cite{allentilde}.
Every set of simulations has been started at a low enough $T^*$ with
a highly ordered configuration where all rods were completely
stretched along the z axis of the laboratory frame of reference 
and arranged in a hexagonal
closed packed fashion.
Simulations at higher values of $T^*$ have been started from an equilibrated
configuration at a lower temperature. For every value of $K$, at least
10 values of temperature have been examined, but for the more 
rigid cases this number has been doubled.

Generally, equilibration run of $10^7$ time--steps have been
performed, followed by as many time--steps of production. The time--step
employed has been 1.36$\times 10^{-4}$$t^*$,
with $t^*$=$\left(m/\epsilon\right)^{1/2}\sigma$.

Additional simulation runs have been performed for certain values of
$K$ and $T^*$ on larger systems of 2400 particles. Transition temperatures
have been observed to remain 
within 5$\%$ of those
determined, as described next, for the smaller systems. 

\begin{figure}
{\par\centering \resizebox*{8.5cm}{!}{\includegraphics{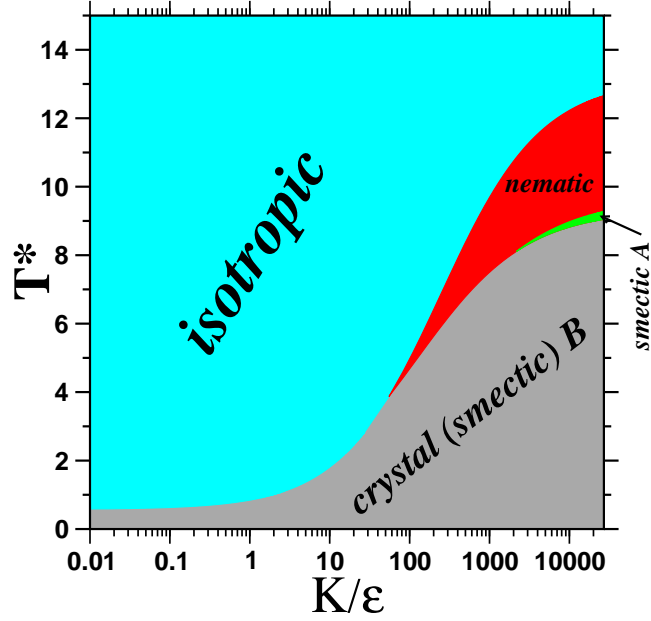}} \par}
\vspace{1.3cm}
{\par\centering \resizebox*{8.5cm}{!}{\includegraphics{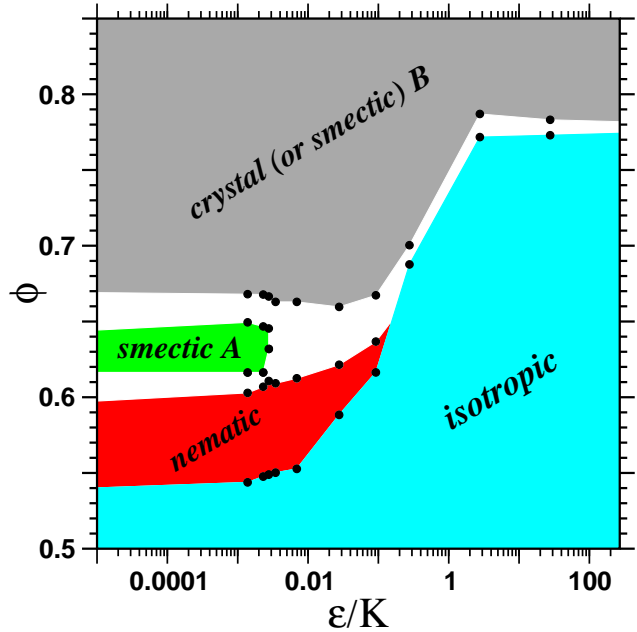}} \par}
\caption{(Color online) Phase Diagram of the semiflexible nine beadlace model as
a function of internal flexibility degree. 
In the top panel the K$/\epsilon$
$versus$ $T^*$ plane is shown, while in the bottom panel the
$\epsilon/K$ $versus$ $\phi$ plane is reported. In both panels,
the regions of existence of the various phases are indicated
by the respective labels,  while the white
region in the bottom panel is the coexistence region.
In the top panel, the transition temperature curves are obtained
by fitting the corresponding computer simulation data with power law functions 
of the type $a\frac{b+K^\alpha}{c+K^\alpha}$.
In the bottom panel, the coexistence data  obtained in the computer simulations
are reported as black dots, and the lines connecting  them are just guides to the
eye.}
\label{figura1}
\end{figure}

Via the calculations of the hexatic ($\Psi_6$), smectic ($\tau$), nematic ($S_2$) order
parameters 
(for the definition of these order parameters
\emph{vide}, \emph{e.g.}, Ref.\cite{prenostro}), four phases have been identified. The 
crystal (or smectic) B phase,
where all order parameters are positive; the smectic A phase, characterized
by $\Psi_6$ equal to, and the other two order parameters larger than, zero; the nematic
phase, with only $S_2$ being non--zero; and the isotropic phase where all parameters 
do, or essentially, vanish.
The resulting phase behavior is depicted in Figure \ref{figura1}, where the sequence
of phases is shown either as a function of $T^*$ and of $\phi$, the packing fraction \cite
{frazioneimpacchettamento}.

Lots of work have been done on the isotropic--to--nematic phase transition
\cite {KS, DijkstraFrenkel, yethiraj}.  
The consensus reached is general. Flexibility increases the density at which
this phase transition occurs.
The results of Fig.\ref{figura1} clearly agree with
this picture.
In common to prior simulations
on short hard spherocylinders \cite{BolhuisFrenkel},
only a single coexistence density is reported in this figure,
as the I--N phase transition in all range of internal flexibility is
so weak that
the two relevant coexistence densities cannot be clearly discerned.

Significant less attention has been dedicated so far to the smectic
phases in systems of semi--flexible rods. The higher density part
of the phase diagram of Fig.1 is therefore of primary concern 
in the present work.

Circa 10 years ago a set of theoretical,
simulational and experimental papers appeared on the
N--S$_A$ phase transition in semiflexible rod systems\cite{fradenfd,
tkachenko,schoot,bladon},  while  more recent theoretical results on this subject
have  been reported in Ref. \cite{sullivan, sullivan2}.

Experiments on suspensions of the long and slightly flexible 
$fd$ virus revealed that the N-S$_A$ phase transition in these
systems is of first--order character and occurring to a packing
fraction of $\sim$ 0.75, while the layer spacing of the 
inhomogeneous phase was very close to the rodlike particle length.
By comparing these results with either numerical data
on systems of perfectly aligned hard rods and the analogous
phase transition in the suspension of TMV virus, it
was speculatively concluded  that flexibility
has three basic effects on the N--S$_{A}$ phase transition:
I) it shifts  the N--S$_A$ coexistence densities to higher values;
II) it drives the transition to being first--order;
III) it  reduces the layer spacing.

Concomitantly, two theories, one phenomenological in character \cite{tkachenko},
the other \cite{schoot} based on an  extension of the Khoklov--Semenov 
theory \cite{KS} to the N--S$_{A}$ phase transition, were developed, both 
adding support to the conjectures coming from the experimental
work. The most recent theory of Ref. \cite{sullivan} is also
confirming these conjectures.
 
The work of Ref. \cite{tkachenko} illustrates the basic mechanism
responsible for the effect of flexibility on the N--S$_{A}$ phase transition
in perfectly aligned rods, originally put forward in Ref. \cite{meyer}.
The formation of a smectic A phase is entropically driven 
by the collective tendency of the rods to fill the voids
present at each end of the particles in the nematic phase.
Although forming layers has a free energy cost, in that
way the rods fill those voids and gain much
free volume in the directions perpendicular to the
layer normal, thus reaching a state of overall larger 
entropy. 
By letting the particles deflect, the above--mentioned
voids can be more efficiently filled already in the nematic
phase by an appropriate bent of the chain. The net result
is that the nematic--smectic phase transition is postponed
to a higher density.

The results of the present work are certainly not in contrast
with  this picture. Indeed, the density at which the nematic
phase remains stable increases with increasing
flexibility. 
In both the oldest theoretical works \cite{tkachenko,schoot},
only spinodal calculations were performed 
so nothing could be said about the actual density
gap characterizing the phase transition, and thus
no  conclusion could be given about the order
of the latter. Binodal points have been instead evaluated
in the work of Ref. \cite{sullivan}, where it is found
that the N--S$_A$ transition is quasi second order in the
perfectly rigid rod case and that flexibility drives the
character of this transition to be first order.
In this work,  
in agreement with previous results on the completely rigid version of the present model
\cite {prenostro}, 
as well as 
on  hard spherocylinder  systems \cite{BolhuisFrenkel, canada}, it is observed that the 
N--S$_A$ transition is actually always first order, being
characterized by a finite density gap and a discontinuity   of
the smectic order parameter $\tau$. The density gap
seems  to slightly decrease by introducing  and increasing 
the internal flexibility.
In fact, while the nematic coexistence line increases with increasing
particle flexibility, that of the smectic A phase remains essentially constant,
at least in the regime where the rods can be
considered only slightly flexibile. 
This fact appears in  agreement with a more recent observation
on the effect of flexibility on the N--S$_A$ phase transition
in $M13$ virus suspensions \cite{PurdyFradenPRE}. In these experiments
the lowest packing fraction at which the smectic A
phase is observed  was found to be  strongly dependent on the
value of ionic strength, in contrast with the previous
results for the $fd$ virus suspensions, but
 essentially independent on the degree of
particle flexibility. For the largest value of ionic strength considered,
for which the rodlike virus
particles may be more closely assimilated to rods interacting
through repulsive and short--range interactions, the above--mentioned
characteristic packing fraction can be estimated to be $circa$ 0.6--0.7,
that is in good agreement with what one can observe in Figure 1.

Thus, the above--mentioned  conjecture about
the effect of flexibility on the order 
of the phase transition is not supported
by the present numerical data. It must be said 
that  this conjecture
was originated  by a  comparison
between systems, like perfectly aligned hard rods
on one hand, and $fd$ virus suspensions on the other
which cannot be  directly linked.
In addition, in the present work, no significant changes
of the value of the layer spacing has been observed upon
introducing and increasing the degree of particle
flexibility, nor there have been significant variations
of the population of rods which, in the smectic A
phase, lie perpendicular to the director and stay between two
layers \cite{bacchettrastra}. The fraction of these rods have been observed to be
always tiny and there are no indication, as suggested
in Ref. \cite{schoot}, that it increases upon letting the
particles be flexible.

The above--mentioned mechanism \cite{tkachenko} describing
the formation of the smectic A phase in rodlike
particle system can be used to justify the observed
unsensitivity of the density at which the S$_A$ appears
for sufficiently stiff rods, and the
subsequent increase of this characteristic density
for more flexible rods.
Smectic A configurations in which rods are completely stretched 
along their contour axis and organized in liquidlike layers
of  sufficiently high densities correspond to a favorable
free energy because the rigid  conformation is that of
minimum energy and the rods already have found
in the smectic structure  a favorable way to   assembly.
It is only when the degree of flexibility is large enough,
and  the nematic configurations progressively correspond
to a more favorable free energy, that the smectic organization
become advantageous only at higher and higher density.

The results of Fig.\ref{figura1} reveal indeed that
for sufficiently flexible rods the smectic A phase disappears
being squeezed out by the nematic phase at lower density and a  
hexatic phase at higher density. These results
are the confirmation of the past statement
often found in the literature that the smectic A phase can
be ousted by other positionally ordered structures
for sufficiently flexible rods.

The present simulations have found that this phase 
has both a layered structure and a in--layer hexagonal arrangement
of the particles, \emph{i.e.} it is of the  crystalline  or
smectic  B type.
In the stiff rod regime, it is worthwhile noticing  that
there is an indication that the enhancement  of internal
flexibility  favors the B phase with respect to the
S$_A$ phase, as the density at which the B  phase remains
stable is slightly 
reduced. 
This can be understood by observing that the principal effect
of introducing and increasing flexibility in stiff rods is
to increase their effective diameters.
For example, for the value of $K$= 5555$\epsilon$ and at a temperature
of $T^*$=8.85, the average length of a rod is only 0.2\% lower
than that corresponding to the fully stretched conformation, 
whereas the average diameter is 10\% larger than that corresponding
to the perfectly rigid rod, as one can appreciate in Figure 2, where
the distribution function, $P(D)$, of the rod
effective diameter, $D$, is shown \cite{ldmedi}.
\begin{figure}
{\par\centering \resizebox*{8.5cm}{!}{\includegraphics{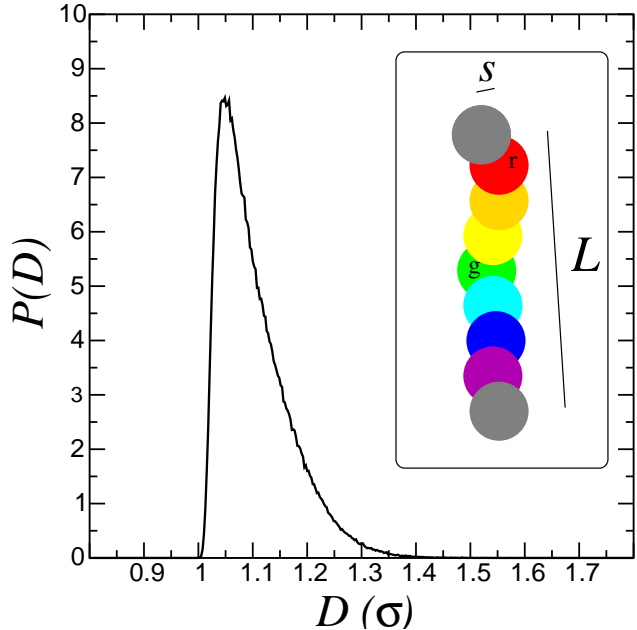}} \par}
\caption{(Color online) Distribution function of the rod effective diameter. The curve refers
to the case of $K$= 5555$\epsilon$  and e
$T^*$=8.85, where the system is in the S$_A$ phase.
The inset is an illustration of the definition of $L$ and $s$,
the two quantities entering the adopted definitions of
effective length and diameter of a wormlike rod \cite{ldmedi}.
Thus, $L$ is the distance
between the two extreme (gray) beads, while $s$, in
the particular case shown, is the distance between the centers of the
r (red) and g (green) beads, resolved along the line joining the 
two gray beads.}
\end{figure}
Thus, the rods are effectively
thicker, so that their projections
onto planes perpendicular to the director
are quasi-two dimensional disks of larger diameters.
This induces, in those that may
be assimilated to a quasi--two dimensional systems of
disks, a crystallization transition at a lower density.
Thus, the smectic A phase results to be bracketed by the
B  and nematic phases until a triple B phase--smectic A--nematic
point  is reached, after which  smectic A phase is no more observed,
and a direct B phase--nematic phase transition occurs.
For smaller values of $K$                  
the density at which the B phase remains stable starts
to  increase. In the same regime the nematic phase interval becomes
thinner until a B phase--nematic--isotropic triple point emerges
after which a direct B phase--isotropic phase transition takes
place.

The phase diagram of Fig.\ref{figura1} is qualitatively similar to 
those resulting from a few theoretical calculations \cite{selinger,herzfeld}. 
In particular, it is in a certain  accord with the phase diagram
presented in Ref.\cite{selinger}, if one identifies their hexagonal
phase  with the B phase described in the present work and 
forget about the smectic A phase, not taken into account
in the theoretical calculations of Ref. \cite{selinger}.
It should be also said that when referring to the hexagonal phase,
the authors of Ref. \cite{selinger} mean  a hexagonal columnar
phase of the type often exhibited by discotic liquid crystals \cite{angewa}
and also observed in suspensions of DNA \cite{livolant}.
However, columnar order 
has been never observed in this work. Perhaps, this is because
the contour length of the rods is not large enough. Computer simulations
on longer rods 
could help clarifying whether internal flexibility could really favor
the formation of columnar phase in long rods or the formation
of this mesophase in the above--mentioned suspensions is due to
other factors, as,  \emph{e.g.},  system polydispersity
\cite{holyst}) and  electrostatic interactions \cite{wensik}.


\begin{thebibliography}{10}
\bibitem{frazioneimpacchettamento} The packing fraction $\phi$ has been calculated
by assuming, as an estimate
of the particle volume,  the value of $\frac{\pi}{4} 5.8 \sigma^3$.
\bibitem{ldmedi} The length of a rod is defined as $\sigma+L$, with $L$
just the distance between the centers
of the two extreme beads. The diameter of a rod is defined as $\sigma+s$, with
$s$ the largest among the distances 
between two any  beads inside a wormlike rod,
projected perpendicularly to the 
axis joining the two extreme beads.
\end{thebibliography}
\end{document}